\documentstyle[prl,floats,aps,twocolumn,epsf]{revtex}

\tighten

\begin{document}
\draft

\twocolumn[\hsize\textwidth\columnwidth\hsize\csname
@twocolumnfalse\endcsname

\title{Universal Self Force from an Extended Object Approach}
\author{Amos Ori and Eran Rosenthal}
\address{Department of Physics, Technion-Israel Institute of Technology, Haifa
32000, Israel}
\date{\today}
\maketitle

\vspace{2ex}

\begin{abstract}

We present a consistent extended-object approach for determining the
self force acting on an accelerating charged particle. In this approach
one considers an extended charged object of finite size $\epsilon $, and
calculates the overall contribution of the mutual electromagnetic forces.
Previous implementations of this approach yielded divergent
terms $\propto 1/\epsilon $ that could not be cured by mass-renormalization.
Here we explain the origin of this problem and fix it. We obtain a consistent,
universal, expression for the extended-object self force, which conforms with
Dirac's well known formula.

\end{abstract}
]

When a charged particle is accelerated in a non-uniform manner, it exerts a
force on itself. This phenomenon of \textit{self force} (often called
''radiation-reaction force'') is known for almost a century, since the
pioneer works by Abraham \cite{Abraham1,Abraham2} and Lorentz \cite{Lorentz}
on the structure of the electron. The non-relativistic form of this force
was obtained by Abraham and Lorentz, who found it to be proportional to the
time-derivative of the acceleration. Later Dirac \cite{Dirac} derived the
covariant relativistic expression for the self force acting on a point-like
particle [Eq. (\ref{fsfinal}) below].

The self-force is a remarkable phenomenon, because essentially it means that
a charged particle may ''exert a force on itself''. A natural approach for comprehending 
this phenomenon within the framework of
classical electrodynamics is the \emph{extended-object approach}. In this
approach one considers an extended charged object of finite size $\epsilon $,
and sum all the mutual electromagnetic forces that its various charge
elements exert on each other. Then one applies the limit $\epsilon
\rightarrow 0$, to obtain the self force in the point-particle limit .
Obviously, if the charged object is static, the mutual forces will always
cancel each other. However, if the charged object accelerates (under the
influence of some external force), one generically finds that the sum of all
mutual forces does not vanish. One would naturally be tempted to identify
this nonvanishing ``total force'' as the self force acting on the particle.
There is a problem, though: The resultant expression obtained for the
``total force'' usually includes a term that diverges like $1/\epsilon $.
This divergent term must somehow be eliminated in order to obtain a
physically meaningful notion of self-force. One would be tempted to apply
the mass-renormalization procedure for this goal. In this procedure one
re-defines the particle's rest mass so as to include the electrostatic
energy $E_{es}\propto \epsilon ^{-1}$. This effectively adds a term $
E_{es}a^{\mu }$ to the total force, where $a^{\mu }$ is the particle's
four-acceleration (we use $c=1$ and signature $(-+++)$
throughout). Unfortunately the $O(1/\epsilon )$ term obtained in previous analyses
was found to depend on the object's shape \cite{GriffithsOwen}, and
generally it does not have the form $-E_{es}a^{\mu }$ that would allow its
elimination by mass-renormalization. In the special case of a charged
spherical shell, Lorentz obtained an overall mutual force that diverges like
$-(4/3)E_{es}a^{\mu }$, which is $4/3$ times larger than what required.
Several authors later confirmed the presence of this problematic $4/3$
factor in spherically-symmetric charge distributions \cite
{GriffithsOwen,Schott,PanofPhil,Fermicomment}.\footnote{One should not
confuse this with another ''$4/3$ problem'', namely the ratio
between the momentum and energy of the electromagnetic field of a
slowly-moving charged particle. These are two distinct (though perhaps
related) problems; see \cite{GriffithsOwen}. Poincare \cite{Poincare}
proposed a solution to the second problem, which involves the
non-electromagnetic internal stresses supporting the charged object. These
short-range internal stresses do not contribute, however, to the overall
mutual force (they merely contribute to the bare mass).} For non-spherical
configurations the situation was found to be even worse \cite{GriffithsOwen}:
in this case the divergent $O(1/\epsilon )$ term is not even co-directed
with $a^{\mu }$, thereby rendering the mass-renormalization procedure
totally inapplicable.

In this manuscript we show that the problem described above stemmed from
applying a too naive notion of ``total force'' (namely, a too naive
summation scheme for the mutual forces). Based on energy-momentum
conservation and proper relativistic kinematics, we formulate the correct
method of summation. By applying this summation method to the mutual forces
we obtain a universal (i.e. shape-independent) $O(1/\epsilon )$ term, which
is precisely of the form $-E_{es}a^{\mu }$, so it is fully annihilated by
mass-renormalization. This resolves the $4/3$ problem, as well as the more
general, more severe (but less well known) directionality problem associated
with the $O(1/\epsilon )$ term in non-spherical configurations. Then from
the remaining $O(\epsilon ^{0})$ term we obtain a universal expression for
the self force (as $\epsilon \rightarrow 0$), which coincides with Dirac's
\cite{Dirac} well known formula. The extended-object approach thus provides
a simple and consistent interpretation of the self force in terms of
standard, non-singular, classical electrodynamics.

We first study an elementary configuration of a ''dumbbell'' \cite
{GriffithsOwen} which consists of two point charges situated at the edges of
a short rod of fixed length $2\epsilon $. This analysis is then
naturally generalized to include any charge distribution. The two electric
charges are denoted $q_{+}$ and $q_{-}$ (the subscripts ''+''and ''-'' are
used throughout to denote quantities associated with the two dumbbell's
edges).

Consider first the dumbbell kinematics. The dumbbell's motion is represented
by its central point, whose proper time and worldline are denoted by $\tau $
and $z^{\mu }(\tau )$, respectively. The four-velocity and four-acceleration
of the central worldline are defined by $u^{\mu }\equiv \dot{z}^{\mu }$ and $%
a^{\mu }\equiv \dot{u}^{\mu }$, respectively, where an overdot denotes
differentiation with respect to $\tau $. (We allow here an arbitrary
acceleration $a^{\mu }(\tau )$, presumably caused by an arbitrary external
force $f_{ext}^{\mu }$ acting on the dumbbell.) At any given moment $\tau $
the rod's edges are located at spacetime's points $z_{\pm }^{\mu }(\tau )$,
given by
\begin{equation}
z_{\pm }^{\mu }(\tau )=z^{\mu }(\tau )\pm \epsilon w^{\mu }(\tau )\ ,
\label{zpm}
\end{equation}
where $w^{\mu }(\tau )$ is a unit spatial non-rotating vector (namely, $%
w^{\mu }$ satisfies $w_{\mu }w^{\mu }=1$, $w^{\mu }u_{\mu }=0$, as well as
the Fermi-Walker transport equation, see e.g. \cite{MTW}).

We denote the proper times along the worldlines of the dumbbell's two edges
by $\tau _{\pm }$. The corresponding four-velocities and four-accelerations
are $u_{\pm }^{\mu }\equiv dz_{\pm }^{\mu }/d\tau _{\pm }$ and $a_{\pm
}^{\mu }=du_{\pm }^{\mu }/d\tau _{\pm }$, respectively. A straightforward
calculation yields \cite{OriRosenthal}
\begin{equation}
u_{\pm }^{\mu }=u^{\mu }\,,\,\,\frac{d\tau _{\pm }}{d\tau }=1\pm \epsilon
a_{||}\,,\,\text{\thinspace }a_{\pm }^{\mu }=\frac{a^{\mu }}{1\pm \epsilon
a_{||}}\ ,  \label{kinematics}
\end{equation}
where $a_{||}\equiv a_{\lambda }w^{\lambda }$ is the projection of $a^{\mu }$
on the rod's direction. The first of these equalities implies that, in the
rest frame of the central point, the dumbbell's edges (and similarly any
other dumbbell's point) are at rest, signifying this frame as the rest frame
of the entire dumbbell.

Consider next the mutual electromagnetic forces between the two charges.
Each charge feels a Lorentz force
\[
f_{\pm }^{\mu }=q_{\pm }F_{\pm }^{\mu \nu }u_{\nu }\,,
\]
where $F_{\pm }^{\mu \nu }$ is the retarded electromagnetic-field tensor produced by the other charge $q_{\mp }$, evaluated at $z_{\pm }^{\mu }(\tau )$. To
evaluate these forces we use a local expansion, derived by Dirac \cite{Dirac}
for the retarded electromagnetic field near a point charge. This expansion,
combined with the kinematical relations\ (\ref{kinematics}) yields the
following expression valid up to order $\epsilon ^{0}$:
\begin{equation}
f_{\pm }^{\mu }\cong q_{+}q_{-}\left[ \pm \frac{w^{\mu }}{4\epsilon ^{2}}-%
\frac{a^{\mu }+w^{\mu }a_{||}}{4\epsilon }+\frac{2}{3}(\dot{a}^{\mu
}-a^{2}u^{\mu })\pm Z^{\mu }\right] \,.  \label{mutf}
\end{equation}
Here $Z^{\mu }$ is a certain $O(\epsilon^0)$ quantity (which is the
same for both charges), whose explicit form is not required here as it
always cancels out upon summation. Throughout this paper, the
symbol ``$\cong$'' represents equality up to terms that vanish as
$\epsilon \rightarrow 0$.

In the standard approach (see e.g. \cite{GriffithsOwen}) one simply sums the
two mutual forces to obtain $f_{sum}^{\mu }$:
\begin{eqnarray}
f_{sum}^{\mu } &\equiv &f_{+}^{\mu }+f_{-}^{\mu }  \nonumber \\
&\cong &-\frac{q_{+}q_{-}}{2\epsilon }(a^{\mu }+w^{\mu }a_{||})+\frac{4}{3}%
q_{+}q_{-}(\dot{a}^{\mu }-a^{2}u^{\mu })\,.  \label{fsum}
\end{eqnarray}
The $O(\epsilon ^{-1})$ term of this quantity suffers from the serious
problem indicated above. The part proportional to $a^{\mu }$ is well
understood, but the second part proportional to $w^{\mu }a_{||}$ is
problematic. This second part is directed along the rod, not in the
direction of $a^{\mu }$, so it cannot be annihilated by
mass-renormalization. When integrated over a spherical shell, this second
part yields $-(1/3)E_{es}a^{\mu }$ \cite{GriffithsOwen}, which may be
recognized as the origin of the ''$4/3$ problem''.

\begin{figure}[p]
\leftline{\epsfbox{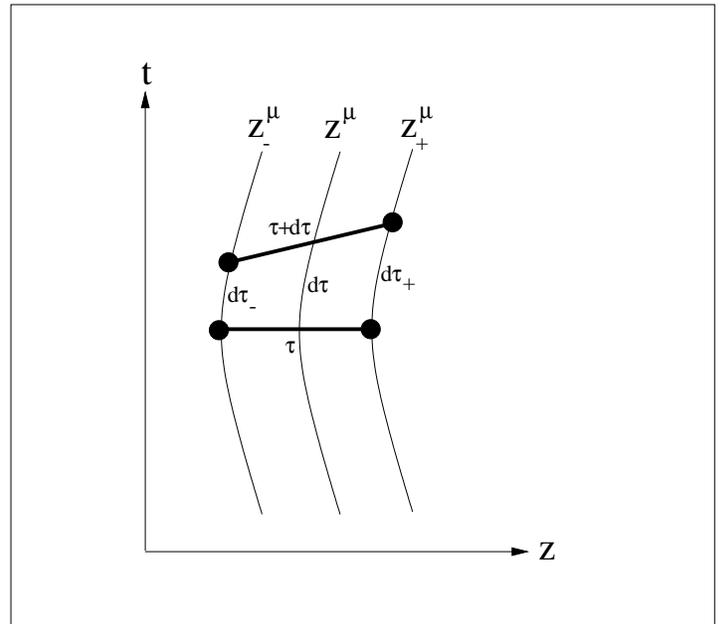}}
\caption{
A spacetime diagram describing the dumbbell's
kinematics. t is the time coordinate (in some inertial reference frame),
and z schematically represents a spatial coordinate.
The dumbbell is represented by a straight bold line,
with the black dots representing the two edge points $z_\pm^\mu$.
Two such bold lines are shown, representing the dumbbell's location in
spacetime at two moments separated by an infinitesimal time interval $d\tau$. The three thin solid lines are
the worldlines of the central point $z^\mu$ and the two edge points
$z_\pm^\mu$.
\label{fig1}}
\end{figure}

This problematic $O(\epsilon ^{-1})$ term indicates that something is wrong
in the identification of $f_{sum}^{\mu }$ with the ''overall mutual
electromagnetic force''. We shall now apply energy-momentum considerations
to resolve this puzzle. Let us denote the total dumbbell's
non-electromagnetic four-momentum, at a given moment $\tau $, by $p^{\mu
}(\tau )$. This momentum can be expressed as an integral of the dumbbell's
stress-energy tensor (not including the electromagnetic stress-energy) over
the dumbbell's momentary rest-frame. From energy-momentum conservation, $%
p^{\mu }$ can change only due to external forces acting on the dumbbell, and
due to energy-momentum exchange with the electromagnetic field. Let us denote
by $dp_{mut}^{\mu }$ the contribution of the mutual electromagnetic forces to this
change in $p^{\mu }$, during an infinitesimal time interval $d\tau $. 
Since the four-momentum is a conserved additive
quantity, we may write $dp_{mut}^{\mu }$ as the sum of the contributions of
the two charges. The contribution coming from the $\pm $ charge to $%
dp_{mut}^{\mu }$ is simply the mutual force $f_{\pm }^{\mu }$ acting on this
charge, multiplied by the proper time $d\tau _{\pm }$ lapsed by this charge 
(between the two ''moments'' $\tau $ and $\tau +d\tau $ ; see figure \ref{fig1})\cite{comment}. Namely,
\begin{equation}
dp_{mut}^{\mu }=f_{+}^{\mu }d\tau _{+}+f_{-}^{\mu }d\tau _{-}\,=\left[
f_{+}^{\mu }\frac{d\tau _{+}}{d\tau }+f_{-}^{\mu }\frac{d\tau _{-}}{d\tau }%
\right] d\tau \,.  \label{dpt}
\end{equation}
We can now identify the term in squared brackets as the ''overall
electromagnetic mutual force'', which we denote $f_{mut}^{\mu }$. Note that
although the two quantities $d\tau _{\pm }/d\tau $ differ from unity (and from
each other) only by an $O(\epsilon )$ quantity, they multiply the large
quantities $f_{\pm }^{\mu }\propto O(\epsilon ^{-2})$; hence the difference
between $f_{mut}^{\mu }$ and $f_{sum}^{\mu }$ may be of order $\epsilon ^{-1}$%
. Indeed a straightforward calculation based on Eqs. (\ref{kinematics},\ref
{mutf},\ref{dpt}) yields
\begin{equation}
f_{mut}^{\mu }\cong -E_{es}a^{\mu }+\frac{4}{3}q_{+}q_{-}(\dot{a}^{\mu
}-a^{2}u^{\mu })\,\,,  \label{fmut}
\end{equation}
where $E_{es}\equiv q_{+}q_{-}/2\epsilon $ is the dumbbell's electrostatic
energy. Note that the $O(\epsilon ^{-1})$ term of $
f_{mut}^{\mu }$ has precisely the right form so as to be cured by mass
renormalization, as we shortly describe.

The remaining $O(\epsilon ^{0})$ term appears somewhat problematic at first glance: It is
proportional to the product $q_{+}q_{-}$, whereas from basic considerations
the self force (which one would like to obtain from $f_{mut}^{\mu }$, after
mass-renormalization, at the limit $\epsilon \rightarrow 0$) should be
proportional to the square of the total charge $q\equiv q_{+}+q_{-}$. This
apparent inconsistency is resolved by noting that the mutual-forces
contribution $f_{mut}^{\mu }d\tau $ is {\em not} the entire momentum exchange with
the electromagnetic field: In addition to the mutual forces $f_{\pm }^{\mu }$
, each charge $q_{\pm }$ also feels its own {\em self force}, which we
denote $\hat{f}_{\pm }^{\mu }$. We shall refer to $\hat{f}_{\pm }^{\mu }$ as
the {\em partial self forces} (to distinguish them from the overall self
force acting on the dumbbell). Obviously it would be inconsistent to ignore
the partial self forces, since our analysis yields a non-vanishing overall
electromagnetic force at the limit $\epsilon \rightarrow 0$ (and by
universality considerations, this result should also apply to the individual
charges $q_{\pm }$). The overall electromagnetic force acting on the
dumbbell, to which we shall refer as the ''bare self force'' $f_{bare}^{\mu }
$, is thus the sum of $f_{mut}^{\mu }$ and the partial self forces:

\[
f_{bare}^{\mu }\cong \left[ -E_{es}a^{\mu }+\frac{4}{3}q_{+}q_{-}(\dot{a}%
^{\mu }-a^{2}u^{\mu })\right] +(\hat{f}_{+}^{\mu }+\hat{f}_{-}^{\mu })\,.
\]
(Although the two new quantities $\hat{f}_{\pm }^{\mu }$ are
apriori unknown, later we shall use a simple argument to relate them to the
overall dumbbell self force, which will allow us to factor them out.)

We now implement the mass-renormalization procedure: We start from the
dumbbell's ``bare'' equation of motion $m_{bare}a^{\mu }=f^{\mu }$, where $f^{\mu
}=f_{ext}^{\mu }+f_{bare}^{\mu }$ is the total (''bare'') force acting on
the dumbbell, and $m_{bare}$ represents the dumbbell's ''bare mass'', i.e.
the total dumbbell's non-electromagnetic energy (in the momentary rest
frame). We define the ''renormalized mass'' $m_{ren}\equiv m_{bare}+E_{es}$.
The equation of motion now takes the form $m_{ren}a^{\mu }=f_{self}^{\mu
}+f_{ext}^{\mu }$, where
\[
f_{self}^{\mu }\equiv f_{bare}^{\mu }+E_{es}a^{\mu }
\]
is the ''renormalized self force''. Note that $f_{self}^{\mu }$ has no $%
O(\epsilon ^{-1})$ term, so we can now safely take the limit $\epsilon
\rightarrow 0$ (after which the approximate equality becomes a precise one).
We find
\begin{equation}
f_{self}^{\mu }=\frac{4}{3}q_{+}q_{-}(\dot{a}^{\mu }-a^{2}u^{\mu })+(\hat{f}%
_{+}^{\mu }+\hat{f}_{-}^{\mu })\,.  \label{fself}
\end{equation}

Consider next the relation between $f_{self}^{\mu }$ and $\hat{f}_{\pm %
}^{\mu }$. Since the self force is the force that a charge experiences due
to its own field, it must scale (for a prescribed worldline) like the square
of the particle's charge. Therefore, the above three self-forces must be
related by $\hat{f}_{\pm }^{\mu }=(q_{\pm }^{2}/q^{2})f_{self}^{\mu }$.
Subtracting $\hat{f}_{+}^{\mu }+\hat{f}_{-}^{\mu }$ from both sides of Eq. (%
\ref{fself}), and noting that
\[
f_{self}^{\mu }-(\hat{f}_{+}^{\mu }+\hat{f}_{-}^{\mu
})=(2q_{+}q_{-}/q^{2})f_{self}^{\mu }\,,
\]
we finally obtain the desired expression for the self force:
\begin{equation}
f_{self}^{\mu }=\frac{2}{3}q^{2}(\dot{a}^{\mu }-a^{2}u^{\mu })\,\,.
\label{fsfinal}
\end{equation}
This agrees with Dirac's expression \cite{Dirac}, and unlike the $O(\epsilon
^{0})$ term in Eq. (\ref{fmut}) it is independent of the dumbbell's charge
distribution (it only depends on the total charge $q$).

The various elements of the above construction of $f_{self}^{\mu }$ can be
summarized by a single mathematical expression:
\begin{equation}
f_{self}^{\mu }=\frac{q^{2}}{2q_{+}q_{-}}\lim_{\epsilon \to 0}\left[
(1+\epsilon a_{||})f_{+}^{\mu }+(1-\epsilon a_{||})f_{-}^{\mu }+\frac{%
q_{+}q_{-}}{2\epsilon }a^{\mu }\right] \,,  \label{final}
\end{equation}
whose all elements have clear meaning and justification, as discussed above
(recall $d\tau _{\pm }/d\tau =1\pm \epsilon a_{||}$).

The above analysis can easily be generalized to include a general charge
distribution: either an extended object consisting of $N$ point charges, or
a continuous charge distribution. (The full analysis will be given elsewhere \cite{OriRosenthal}). Essentially one needs to sum over the contributions of each pair
of charges (or charged volume elements) to the overall mutual force; and for
each such pair, the contribution is given by the above dumbbell-model
analysis. In the continuous case there is no need to consider the ``partial
self forces'' as their contribution vanishes. (This can easily be seen from the limit 
$N\rightarrow \infty $ of the discrete model, in which the individual
charges scale like $1/N$, and correspondingly the partial self forces scale like $%
1/N^{2}$.) In both the discrete and continuous cases, we obtain the result (%
\ref{fsfinal}), with $q$ being the total charge.

We conclude that at the limit $\epsilon \rightarrow 0$ the ``total
electromagnetic force'' acting on any extended charged object is {\em %
universal}, which provides a simple interpretation to the notion of self
force.

%************************************************************************

%\begin{references}

\begin{references}
\bibitem{comment}
After this work was completed, we became aware of an
apparently related approach that was applied by Pearle to the case of
spherical shell, in order to overcome the ''$4/3$ problem'' \cite {Pearle}. Our analysis, however, is simpler, more transparent, and much more general, as it applies to any charge distribution. In particular, it solves the more severe directionality problem in the non-spherical case.
\end{references}
\end{document}